\documentstyle[twoside,fleqn,aip2col]{article}
\newcommand{\AmS}{{\protect\the\textfont2
  A\kern-.1667em\lower.5ex\hbox{M}\kern-.125emS}}

\input epsf
\newcommand{\psfile}[1]{
  \setlength{\epsfxsize}{\columnwidth}
  \epsfbox{#1}\vspace{-15pt}
}

\def\journal#1&#2(#3){\begingroup \let\journal=\dummyj@urnal
    \unskip, \sl #1\unskip~\bf\ignorespaces #2\rm
    (\afterassignment\j@ur \count255=#3), \endgroup\ignorespaces }
\def\j@ur{\ifnum\count255<100 \advance\count255 by 1900 \fi
          \number\count255 }
\def\dummyj@urnal{%
    \toks@={Reference foul up: nested \journal macros}%
    \errhelp={You forgot & or ( ) after the last \journal}%
    \errmessage{\the\toks@ }}
\def\apsjournal#1&#2,#3(#4){\unskip, #1\ {\bf #2}, #3 (19#4)\unskip}
\def\prl{\apsjournal Phys.\ Rev.\ Lett.}
\def\prd{\apsjournal Phys.\ Rev.\ D}
\def\pr{\apsjournal Physics\ Reports}
\def\pl{\apsjournal Phys.\ Lett.}

\def\*{\hbox{$\rm *$}}

\def\fb{\rm{fb}$^{-1}$}

\def\beqa{\begin{eqnarray}}
\def\eeqa{\end{eqnarray}}
\def\nn{\nonumber}
\def\tbar{\bar{t}}
\def\qbar{\bar{q}}

\title{Top Quark, W-boson and Light Higgs}
 
\author{Stephen Parke
\address {Department of Theoretical Physics\\
Fermi National Accelerator Laboratory\\
Batavia, IL 60500-0500\\
U.S.A.}
\thanks{
Presented at the First International Lecture and Workshop Series
Frontiers in Contemporary Physics; ``Fundamental Particles and 
Interactions,'' May 11-16, 1997 at Vanderbilt University, Nashville TN.
}
}
       
\begin{document}
 
\begin{abstract}

In this presentation I will concentrate on top quark physics, 
the W-boson mass and the possibility of discovering 
a light Higgs boson via associated production at the Fermilab Tevatron.

\end{abstract}
 
\maketitle
 
\section{INTRODUCTION}
The top quark, the W-boson and the Higg boson form an 
interesting triptych of elementary particles.
In the Standard Model knowing the mass of two of these particles,
usually the top quark and W-boson, we can
predict the mass of the third, the Higgs boson.
Therefore in this proceedings I will primarily cover the following topics,
top quark physics, W-boson mass and the light Higgs boson at the
proton-antiproton collider at Fermilab, the Tevatron. 
Other hadron collider topics to be cover in this conference include
B-physics \cite{Butler},
QCD \cite{Bergerqcd},
Electroweak Physics \cite{Marciano} 
and
Supersymmetry \cite{Barger}.

\section{TOP QUARK PHYSICS}
The most surprising thing about the top quark is that its mass is approximately
175 GeV, nearly twice as heavy as the W and Z bosons and more than 30 times 
the mass of its electro-weak partner the b-quark. 
The Yukawa coupling constant of the top quark
\beqa
\quad \quad \quad m_t \sqrt{2\sqrt{2}G_F} & \sim & 1 \nn
\eeqa
whereas for the electron the Yukawa coupling is $3 \times 10^{-6}$.
Why is the top quark so heavy? 
Does it have a special roll in Electro-Weak
symmetry breaking?
Does it have Standard Model couplings? 
These are some of the question that need to be answered.

\subsection{Pair Production}
At a hadron collider the dominant mode of top quark 
production at hadron colliders is via 
quark-antiquark annihilation or gluon-gluon fusion
\beqa
\quad \quad q \qbar & \rightarrow & t \tbar \nn \\
g g     & \rightarrow & t \tbar. \nn
\eeqa
\begin{figure}[hbt]
  \psfile{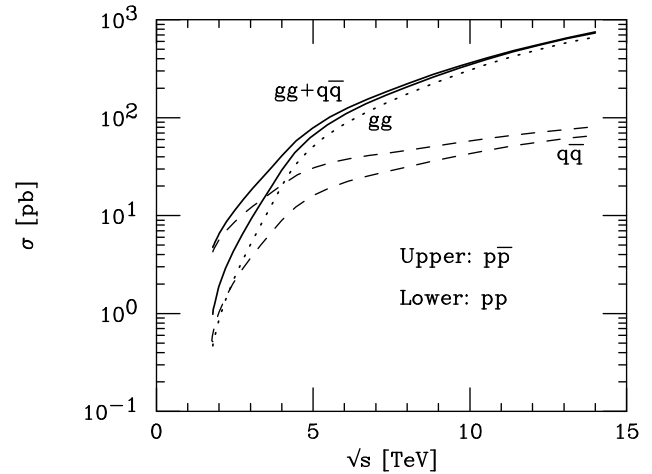}
  \caption{\label{fig-roots} 
The lowest order QCD top quark pair production cross sections 
as a function of $\sqrt{s}$ for a 175 GeV top quark mass.
}
\end{figure}
\begin{figure}[hbt]
  \psfile{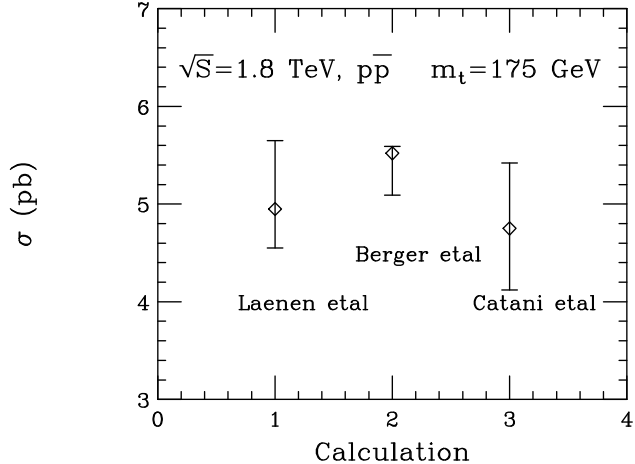}
  \caption{\label{fig-all} 
Resummed Next to Leading Order top quark cross sections from
Laenen etal$^{~\cite{Laenen}}$, Berger etal$^{~\cite{Berger}}$
and Catani etal$^{~\cite{Catani}}$.
}
\end{figure}

Fig~\ref{fig-roots} is the lowest order cross sections for 
these subprocesses verses $\sqrt{s}$ for
$ m_t ~=~ 175~GeV $ for both proton-antiproton and proton-proton
accelerators.
For the Tevatron the dominant production mechanism,
80 to 90 \% of the total cross section,
is quark-antiquark annihilation 
whereas at the LHC gluon-gluon fusion is 80 to 90 \% of the total.
At the Tevatron the top quark pairs are produced with a typical
speed in the zero momentum frame of 0.6c whereas at the LHC this
speed is 0.8c.

Recently a number of authors$^{~\cite{Laenen}~ -~\cite{Catani}}$
have calculated the cross section for top
quark pair production not only at next to leading order but they have 
summed the large logarithms to all orders in perturbation theory.
For the Tevatron these results are displayed in Fig~\ref{fig-all}.
Even though these authors all agree on the top cross-section at the Tevatron
they disagree in principal on how these calculations should be performed.

\begin{figure}[htb]
  \psfile{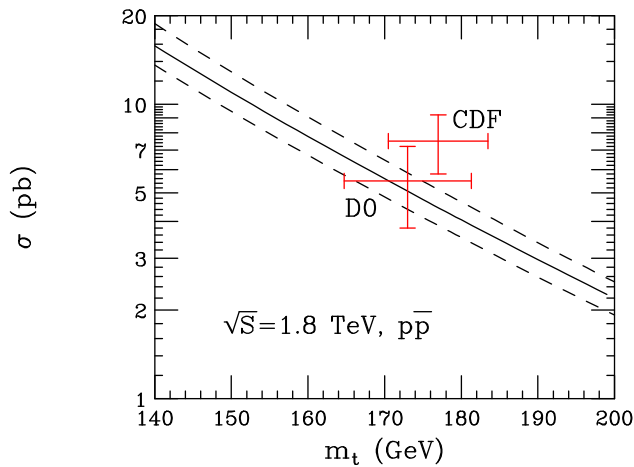}
  \caption{\label{fig-xsec} 
The dependence of the top quark cross section as a function of the 
top quark mass for the Catani et al calculation \cite{Catani}. 
The latest experimental results are also shown.
}
\end{figure}

Fig~\ref{fig-xsec} is the cross section verses the mass of the top
quark for the calculation by Catani et al \cite{Catani}. The functional 
dependence of the other calculations is essentially the same
with the cross section dropping by a factor of 2 for every 20 GeV increase
in the top quark mass.
Also shown on this figure are the results from CDF and D0. 

In raising the energy of the Tevatron from 1.8 to 2.0 TeV
the top cross section increases by 38 \% with the gluon-gluon fusion component
increasing from 10 to 20 \% of the total.

\subsection{Top Quark Decay}
\begin{figure}[htb]
  \psfile{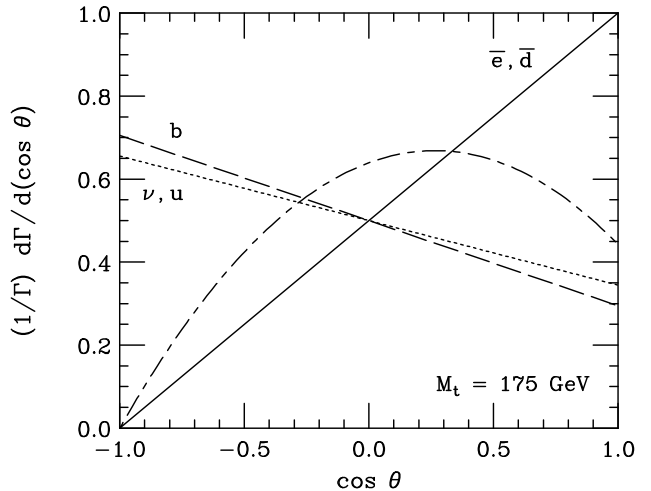}
  \caption{\label{fig-decay-corr} 
Angular correlations in the decay of a 175 GeV spin-up top quark. 
The labeled lines are the angle between the spin axis and the particle in the
rest frame of the top quark. 
The unlabeled dot-dash line is the angle between the b quark and the positron
(or d-type quark) in the rest frame of the W-boson.
}
\end{figure}

\begin{figure}[hbt]
  \psfile{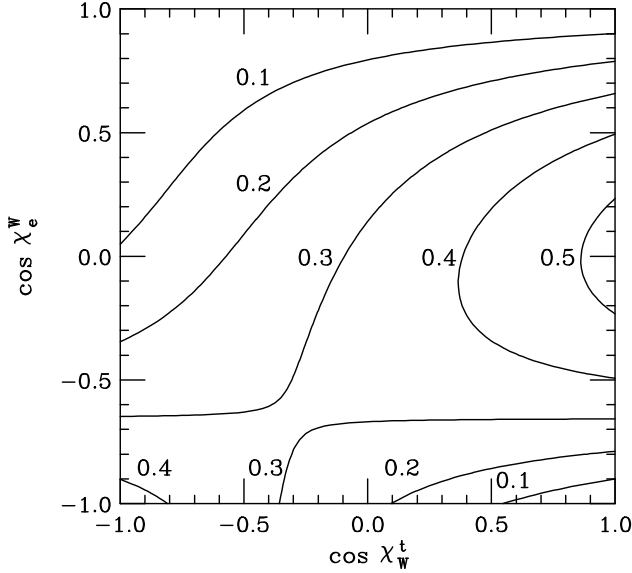}
  \caption{\label{fig-contours} 
Decay distribution contours.
$\chi^t_W$ is the angle between the top-quark spin 
and direction of motion if the W-boson in the top quark rest frame.
$\pi-\chi^W_e$ is the angle between the direction of motion of the 
b-quark and the positron in the W-boson rest frame.
}
\end{figure}
In the Standard Model the top quark decays primarily 
into b-quark and a W boson,
\beqa
\quad \quad t \rightarrow  b ~W^+ & \rightarrow & b ~l^+ ~\nu \nn \\
& \rightarrow & b ~\bar{d} ~u. \nn
\eeqa
For a 175 GeV the width of this decay mode is 1.5 GeV, 
see Bigi etal$^{\cite{Bigi}}$. 	   
Thus the top quark decays before it hadronizes and any spin information
introduced in the production mechanism is passed on to the decay products.
Fig~\ref{fig-decay-corr} gives the correlations of the decay products
with the spin direction for a polarized top quark$^{\cite{Jezabek}}$. 
Also shown on this figure is the correlation of the charged lepton (or d-type
quark) with the b-quark direction in the W-boson rest frame showing the
$m_t^2:2m_W^2$ ratio of 
longitudinal to transverse W-bosons in top quark decay.
Fig~\ref{fig-contours} shows the correlations between the W-boson decay 
direction relative to the spin-direction and the charge lepton 
(or d-type quark)
direction relative to the minus b-quark direction in the W-boson rest frame.
That is, if the $W^+$ is emitted in the spin direction it is longitudinal
and in the minus spin direction it is transverse.

\subsection{Spin Correlations in Pair Production}
\begin{figure}[hbt]
  \psfile{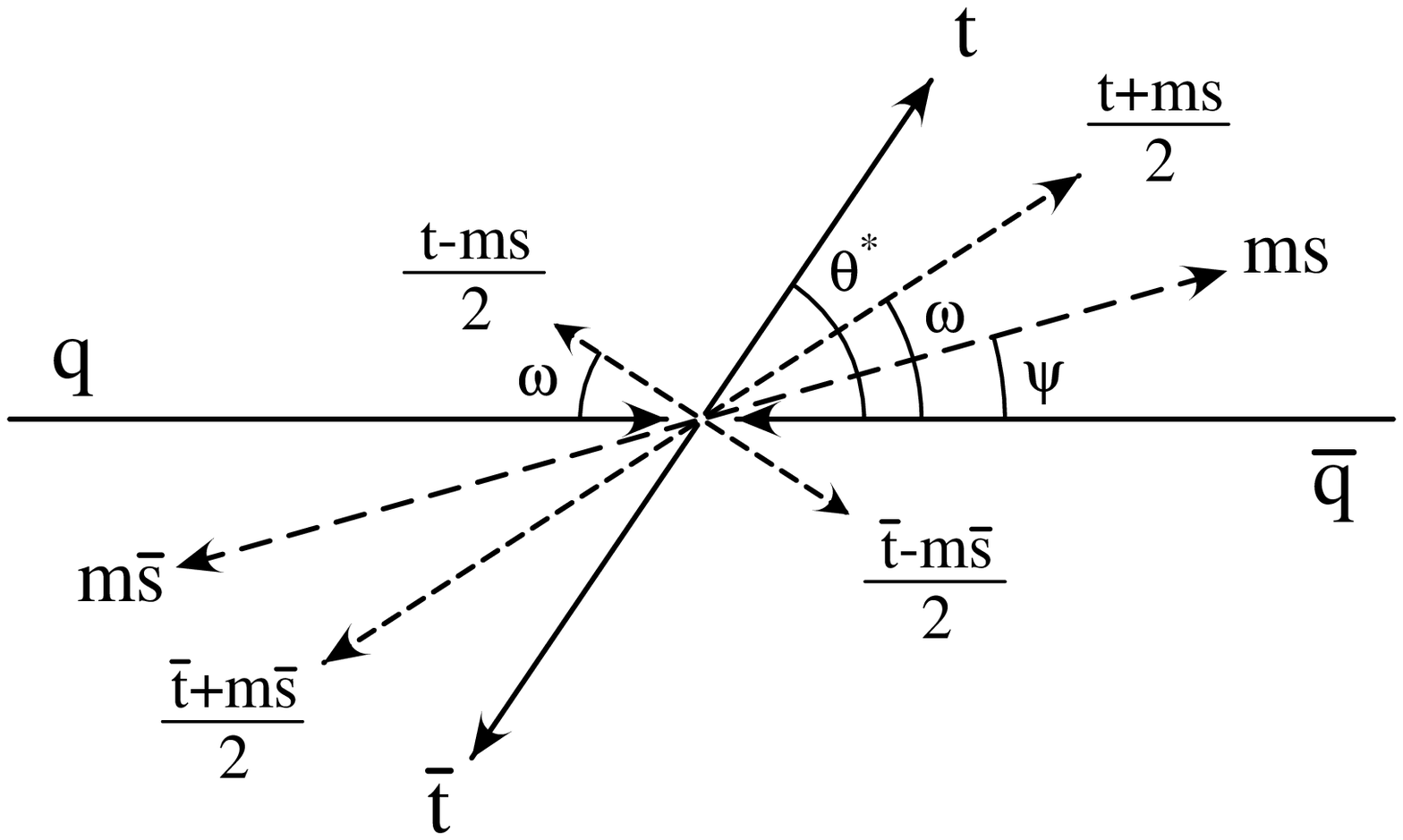}
  \caption{\label{fig-spins} Spin and Momentum vectors 
for $q\qbar \rightarrow t\tbar$} in the zero momentum frame.
\end{figure}

\begin{figure}[hbt]
  \psfile{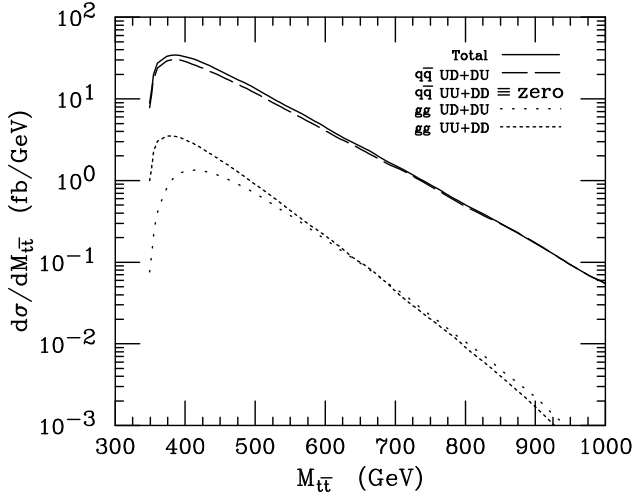}
  \caption{\label{fig-mtt} Invariant mass distribution of the $t\tbar$ pair
decomposed into the spin components Up-Down, Down-Up, Up-Up and Down-Down
using the Off-Diagonal spin basis.
}
\end{figure}
Fig~\ref{fig-spins}  is the relevant three vectors for the 
spin correlation studies of the top quark pairs
produced by quark-antiquark annihilation$^{\cite{MPSW}}$.
If the angle $\psi$ is chosen such that 
\beqa
\quad \quad \tan \psi & = & { \beta^2 \cos \theta^* \sin \theta^*
\over
(1-\beta^2 \sin^2\theta^*)} \nn
\eeqa
then the top quarks are produced in only the 
Up-Down and Down-Up configurations {\it i.e.} the Up-Up and Down-Down
components identically vanish$^{\cite{PS},~\cite{MP}}$.
This spin basis is known as the Off-Diagonal basis.

For the Up-Down spin configuration, the preferred emission directions for
the charged leptons (or d-type quarks) of the top and anti-top quark are
given by the directions of $(t+ms)/2$ and $(\tbar+m\bar{s})/2$ respectively.
Whereas for the Down-Up configuration the preferred directions are
$(t-ms)/2$ and $(\tbar-m\bar{s})/2$ respectively.
These vectors make
an angle $\omega$ with respect to the beam direction with
\beqa
\quad \quad \sin \omega & = & \beta \sin \theta^*. \nn
\eeqa
Near threshold $\psi,~\omega \approx 0$ whereas for ultrarelativisitic tops
$\psi, ~\omega \approx \theta^*$ as expected.

\subsection{New Physics in Production}
\begin{figure}[htb]
  \psfile{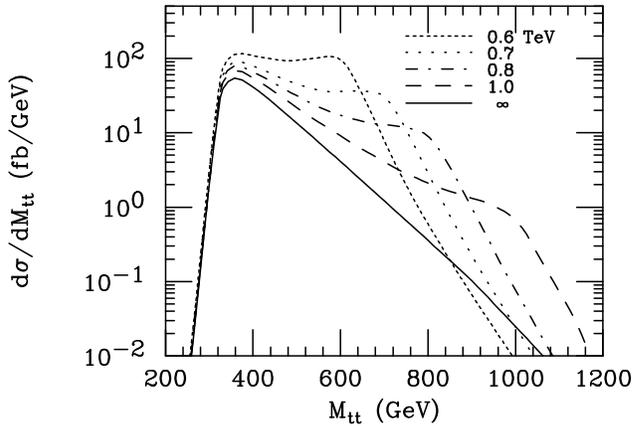}
  \caption{\label{fig-mtttopcolor} 
The invariant mass of the $t\tbar$ pair for the topcolor octet model.}
\end{figure}

Hill and Parke~\cite{HP} have studied the effects of new physics
on top quark production in a general operator formalism as well as in
topcolor models. In these models the distortions in top quark production
and shape are due to new physics in the $q\bar{q}$ subprocess.
The effects of a coloron which couples weakly to the light generations
but strongly to the heavy generation is given in Fig~\ref{fig-mtttopcolor}.

\begin{figure}[htb]
  \psfile{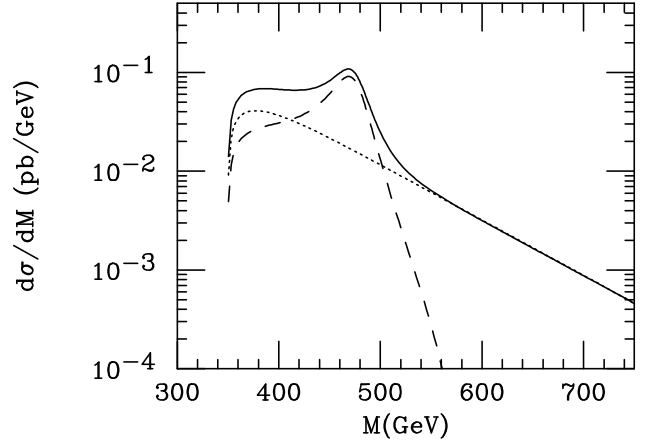}
  \caption{\label{fig-mtttechni} 
The invariant mass of the $t\tbar$ pair for the two scale technicolor model.}
\end{figure}

Similarly Eichten and Lane~\cite{EL} have studied the effects of
multi-scale technicolor on top production through 
the production of a techni-eta
resonance, see Fig~\ref{fig-mtttechni}.
Here the coupling of the techni-eta is to $gg$, therefore only
this subprocess is different than the standard model. 
At the Fermilab Tevatron top production is dominated by $q\bar{q}$ annihilation
while at the LHC it is the $gg$ fusion subprocess that dominates. Therefore
these models predict very different consequences for top production at the 
LHC.

\subsection{Experimental Results}
Both CDF and D0 have observed the top quark pair events in 
the dilepton channel, the lepton plus jet channel and the all jets channel.
To get the latest information on the measurement of the mass 
and production cross sections see the CDF web page at
www-cdf.fnal.gov and the D0 web page at www-d0.fnal.gov.

At the time of this conference the CDF$^{~\cite{CDFTOP}}$ results are
\beqa
\quad \quad  m_t & = & 176.8 \pm 4.4 \pm 4.8~ GeV \nn \\
\sigma_{t\tbar} & = & 7.5^{+1.9}_{-1.6} \pm 1.67 ~pb \nn
\eeqa
and for D0$^{~\cite{D0TOP}}$
\beqa
\quad \quad m_t & = & 173.3 \pm 5.6 \pm 6.2 ~GeV \nn \\
\sigma_{t\tbar} & = & 5.53 \pm 1.67 ~pb \nn
\eeqa
For a comparison with theory see Fig \ref{fig-xsec}.
What is surprising about these results is that with approximately 
100 top quark events in total the top mass is already known quite accurately.
Unfortunately all top quark experimental results so far
are consistent with the Standard Model.

\subsection{Single Top Quark Production}

\begin{figure}[hbt]
  \psfile{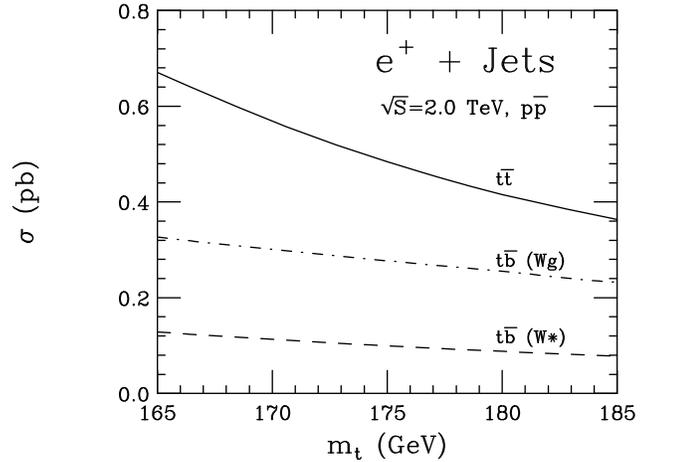}
  \caption{\label{fig-comp} 
The cross sections verse mass of the top quark 
for top pair production$^{~\cite{Catani}}$,
single top via W-gluon fusion$^{~\cite{Stelzer}}$ 
and single top via a virtual W-boson$^{~\cite{Smith}}$
in the channel positron, missing energy plus jets for the 2 TeV Tevatron.
}
\end{figure}

Recently reliable results for the next to leading order calculations
for single top quark production at hadron colliders via
a virtual W-boson$^{~\cite{Smith}}$ or via W-gluon fusion$^{~\cite{Stelzer}}$
have been presented.
A comparison of the rates for these processes
can be found in Fig~\ref{fig-comp} for events with the topology 
positron, missing transverse energy plus jets.  
The rates for both of these single top
processes are proportional to the CKM matrix element $V_{tb}$ squared
therefore these processes can be used to measured this important Standard Model
parameter.
For 2 \fb (30 \fb) of data at the Tevatron the expected 
uncertainty on $V_{tb}$ is 12\% (3\%). 

Single top quark production is a great source of polarized top quarks
with the polarization being in the direction of the d-type quark in the event
{\it i.e.} the anti-proton direction for W$^*$ production and the spectator
jet for production via W-gluon fusion. 
The production of single top quarks
through a virtual W-boson is sensitive to form factors in the
$Wtb$ vertex at a $Q^2 = m_t^2$. 
Hints of new physics could be discovered in this process.

\section{W-BOSON AND LIGHT HIGGS PHYSICS}

In the Standard Model the mass of the top quark, W-boson and Higgs boson
are all related to one another. 
So precision measurements of the
top quark mass and W-boson mass will give us 
information on the Higgs boson mass.

\subsection{W-boson Mass}

\begin{figure}[hbt]
  \psfile{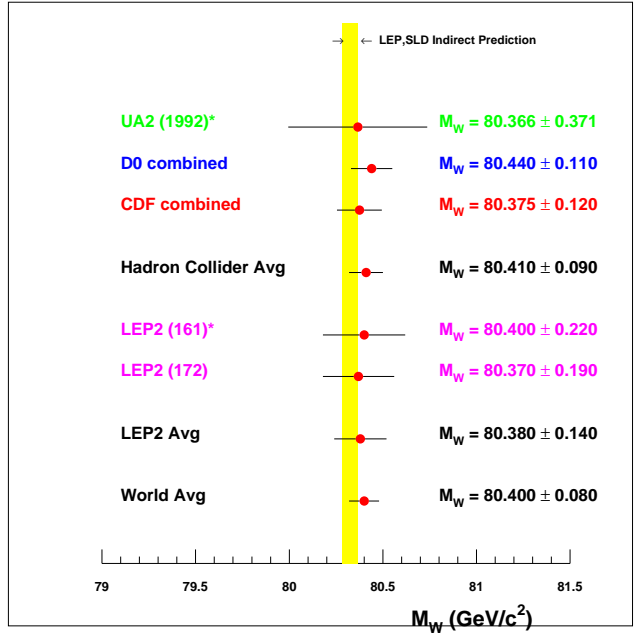}
  \caption{\label{fig-w-mass} 
Summary of W-boson mass measurements.
}
\end{figure}

The latest result on the W-boson mass from CDF$^{~\cite{CDFW}}$ is
\beqa
\quad \quad m_W & = & 80.38 \pm 0.12 ~GeV \nn
\eeqa
and from D0$^{~\cite{D0W}}$ is
\beqa
\quad \quad m_W & = & 80.44 \pm 0.11 ~GeV. \nn
\eeqa
Fig~\ref{fig-w-mass} is a summary of the current results 
from all experiments on the W-boson mass.

\begin{figure}[hbt]
  \psfile{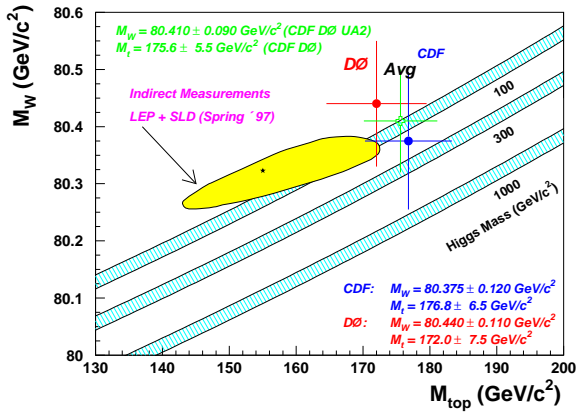}
  \caption{\label{fig-wvt-mass} 
The relationship between the mass of the top quark, W-boson and Higgs
boson in the Standard Model plus the latest experimental information.}
\end{figure}

In the Standard Model, because of radiative corrections,
knowing the W-boson mass and the top quark mass gives a determination of the
Higgs boson mass. 
Fig~\ref{fig-wvt-mass} shows the current experimental results on this indirect 
measurement of the Higgs boson mass.
Unfortunately at the current level of accuracy on the W and top mass
measurements little can be said about the Higgs boson mass except that light
values seem to be preferred.
If this holds up with increased precision on these measurements 
it is great news as it means the Higgs boson is easily accessible 
or there is new physics near at hand.
Either outcome would be great for particle physics.

Improvements in the W-boson mass can be expected from LEP2 possible reaching
an uncertainty of 34 MeV. 
With a few fb$^{-1}$s of data the Tevatron should reach an uncertainty
of 40 MeV on the W-boson mass and almost 2 GeV on the top quark mass.
If TeV33 gets 30 to 100 fb$^{-1}$ of data the uncertainty 
on the W-boson mass will probably reach 20 MeV and top quark mass of 1 GeV.
This would greatly enhance the determination of the Higgs boson mass from 
Fig~\ref{fig-wvt-mass}.


\subsection{Light Higgs Boson}
The mass of the W-boson and mass of the top quark suggest that the Higgs
boson is light. SUSY models predict that the lightest Higgs be less than
150 GeV (125 GeV in the minimal model, MSSM). 
LEP2 will explore up to a mass of 95 GeV by the year 2000.
At the Tevatron a light Higgs$^{~\cite{Stange}}$ can be explored up to a mass of 
130 GeV$^{~\cite{KIM} ~- ~\cite{WOM}}$ with 
sufficient integrated luminosity, 30 - 100 \fb, using the 
subprocess
\beqa
\quad \quad q ~\qbar^{\prime} & \rightarrow & W ~+~ H \nn
\eeqa
with the W decaying leptonically and the Higgs decaying to $b\bar{b}$.
The physics backgrounds for this process are 
\beqa
\quad \quad q ~\qbar^{\prime} & \rightarrow & W ~+~ b ~+~\bar{b} ~~(QCD)\nn \\
		& \rightarrow & t ~+~ \bar{b} \nn \\
		& \rightarrow & W ~+~ Z. \nn
\eeqa
This physics requires double b-tagging with high efficiency and low fake rates.
Also one needs good resolution on the $b\bar{b}$ mass,
above the Z-boson peak, then with very large data sets 
the Higgs boson will be observable if its mass is below 130 GeV.
The process $ q \qbar  \rightarrow  Z ~+~ H $ will also be 
useful$^{~\cite{YAO}}$.
Hopefully the Tevatron can obtain these large data sets 
before the LHC has obtained significant data sets.

\section{CONCLUSION}
Hadron Colliders provide a rich, diverse ``feast of physics''.
The top quark, W-boson and Higgs boson form a very rich triptych
but there is also QCD, B-physics, Electroweak, SUSY etc.
While the Fermilab Tevatron still holds the energy frontier it should be 
exploited to the fullest possible extent with luminosity upgrades to both
accelerator and detectors. 

\section*{ACKNOWLEDGMENTS}
I wish to thank the authors of many  of the references who provided data from their 
work for the plots in this presentation. 
I also thank Professor Panvini and Professor Weiler
of Vanderbilt University for a productive and interesting conference. 

The Fermi National Accelerator Laboratory is 
operated by the Universities Research Association, Inc., under contract
DE-AC02-76CHO3000 with the United States of America Department of Energy.

\end{document}